\documentclass[%
 aip,
 jmp,%
 amsmath,amssymb,
preprint,%
]{revtex4-1}

\usepackage{amsfonts}
\usepackage{bm}%
\usepackage[colorlinks=true,linkcolor=blue,filecolor=blue,menucolor=yellow,urlcolor=blue,
citecolor=blue,anchorcolor=blue]{hyperref}
\usepackage{graphicx}
\usepackage{rotating}
\usepackage{url}
\usepackage{ccaption}
\usepackage{floatrow}
\usepackage[font= large,label font=bf,labelformat=simple]{subfig}

\usepackage[mathlines]{lineno}
\modulolinenumbers[1]
\linenumbers\relax 
\begin{document}
\title{A mathematical model for phenotypic heterogeneity in breast cancer with implications for therapeutic strategies}
\author{Xin Li}
\email{xinlee0@gmail.com}
\author{D. Thirumalai}
\email{dave.thirumalai@gmail.com}
\affiliation{$^1$Department of Chemistry, University of Texas, Austin, TX 78712, USA.}

\date{\today}

\begin{abstract}

Inevitably, almost all cancer patients develop resistance to targeted therapy.  Intratumor heterogeneity (ITH)  is a major cause of drug resistance.  Mathematical models that explain experiments quantitatively  is useful in understanding the origin of ITH, which then could be used to explore scenarios for efficacious therapy. Here, we develop a mathematical model to investigate ITH in breast cancer by exploiting the observation that HER2+ and HER2- cells could divide symmetrically or asymmetrically.    Our predictions for the evolution of cell fractions are in quantitative agreement with single-cell experiments. Remarkably, the colony size of HER2+ cells emerging from a single HER2- cell (or vice versa), which occurs in about four cell doublings,  agrees perfectly with experimental results, without tweaking any  parameter in the model. The theory quantitatively explains experimental data on the responses of breast cancer tumor  under different treatment protocols. We then used  the model to predict that, not only the order of two drugs, but also the treatment period for each drug and the tumor cell plasticity could be manipulated to improve the treatment efficacy. Mathematical models, when integrated with data on patients, make possible exploration of a broad range of parameters readily, which might provide insights in devising effective therapies. 
\end{abstract}


\maketitle

\section*{Introduction}
Nearly 10 million people died of cancer worldwide in 2020\cite{Sung2021}, despite innovations in the development  of many novel drugs.  In principle, the advent of new technologies ought to make  drugs highly efficacious while minimizing toxicity. The next-generation sequencing allows us to design personalized therapy, targeting  specific genetic variants which drive disease progression\cite{ross2014new,ellis2009resistance}.  However, drug resistance ultimately occurs, regardless of  targeted therapeutic protocols, which poses a formidable challenge for oncologists\cite{ahronian2017strategies}. A  deeper understanding of the underlying resistance mechanism could be useful in controlling the tumor burden and its relapse. 

Intratumor heterogeneity (ITH),  which denotes the coexistence of cancer cell subpopulations with different genetic or phenotypic characteristics in a single tumor~\cite{almendro2013cellular,burrell2013causes},  is the prominent cause of drug resistance and recurrence of cancers\cite{marusyk2012intra,dagogo2018tumour,Filho21CancerDiscovery}. With the development of deep-sequencing technologies and  sequencing at the single cell level\cite{koboldt2013next,shapiro2013single}, intratumor {\it genetic} heterogeneity has been observed in many cancer types \cite{gerlinger2012intratumor,vogelstein2013cancer,de2014spatial,navin2011tumour,swanton2012intratumor,li2019share}.  Meanwhile, increasing evidence shows that {\it phenotypic} variations in tumor cells (without clear genetic alterations) also play a crucial role in cancer development,  and is presumed to be one of the major reasons for the development of drug resistance in cancer therapy\cite{meacham2013tumour,marusyk2012intra}. However, the underlying mechanism of ITH induced by the phenotypic variability of cancer cells is still elusive, which represents an obstacle for the development of efficient treatments for cancer patients\cite{altrock2015mathematics}.

The phenotypic heterogeneity of normal cells can emerge from cellular plasticity, which is the ability of a cell to adopt different identities. Cellular plasticity is widespread in multicellular organisms, dictating the development of organism, wound repair and tissue regeneration\cite{moore2006stem,reya2001stem,blanpain2014plasticity}. One of the best known examples is the differentiation hierarchies in stem cells, which leads to the production of progenitor cells, followed by the mature differentiated cells\cite{jopling2011dedifferentiation,gupta2019phenotypic}. 

It has been proposed that cancer might be derived from cancer stem (or initiating) cells (CSCs). The CSCs are similar to normal stem cell, but possess the ability to produce all cell types found in a tumor sample, resulting  in ITH\cite{al2003prospective,clevers2011cancer,kreso2014evolution}. However, the prospects of a hierarchical organization, and also the unidirectional differentiation of CSCs have been challenged by  recent experimental observations\cite{gupta2011stochastic,chaffer2011normal,shackleton2009heterogeneity,quintana2010phenotypic}. Some `differentiated' cancer cells are  capable of switching back to the CSCs in breast cancer \cite{gupta2011stochastic,chaffer2011normal}. Melanoma cells do not show any hierarchically organized structure as cells are capable of switching between different phenotypes reversibly\cite{shackleton2009heterogeneity,quintana2010phenotypic}. Several models that assume reversible state transitions have been proposed to explain the observed stable equilibrium among cancer cell subpopulations with different phenotypes\cite{gupta2011stochastic,zhou2014nonequilibrium}. However, a detailed understanding of the underlying mechanism driving the cell state transition is still lacking, as most previous experimental observations are based on measurements from  bulk cell populations\cite{gupta2011stochastic,chaffer2011normal,quintana2010phenotypic}.

A recent insightful experiment tracked the evolution of a single circulating tumor cell (CTC) derived from estrogen-receptor (ER)-positive/human epidermal growth factor receptor 2 (HER2)-negative (ER+/HER2-) breast cancer patients {\it in vitro}\cite{jordan2016her2}. Surprisingly, HER2+ cells (with expression of HER2) emerge from a cell colony grown from  a single HER2- cell within four cell doublings and vice versa.  The single-cell level experiment demonstrates that  reversible transitions occurred between the two  breast cancer cell types, thus providing a clue to understanding the nature of cancer cell plasticity observed in this and other experiments\cite{gupta2011stochastic,chaffer2011normal,quintana2010phenotypic,jordan2016her2}.  Because normal stem cell can differentiate into non-stem cells through asymmetric cell division\cite{jopling2011dedifferentiation}, it is possible that cancer cells might also change their identity by asymmetric division\cite{granit2018regulation}, which is a potential cause of ITH.

We noticed that the emergence of an altered cell phenotype is to be coupled to cell division, as indicated by the experiments that a cell of a specific genotype produces daughter cells with an altered  phenotype\cite{jordan2016her2}.  Based on this observation, we developed a theoretical model to describe the  establishment of ITH from a single type of breast CTCs. In quantitative agreement with experiments, our model captures the tumor growth dynamics under different initial conditions.  It also naturally explains the emergence and evolution of ITH, initiated from a single cell type, as discovered in a recent experiment\cite{jordan2016her2} .  Without adjusting any free parameter in the model, we predict the evolution of cell fractions and also the colony size for the appearance of HER2+ (HER2-) cell types starting from a single HER2- (HER2+) cell. Remarkably, the predictions agree perfectly with the experimental observations.   As a consequence of ITH, drug resistance develops rapidly, which we also reproduce quantitatively. By exploring a range of parameters in the mathematical model, we found that several factors  strongly influence the growth dynamics of the tumor. The insights from our study may be useful in devising effective therapies\cite{sharma2010chromatin,jordan2016her2}.

\section*{Results}

\begin{figure}
{\includegraphics[width=1.0\textwidth] {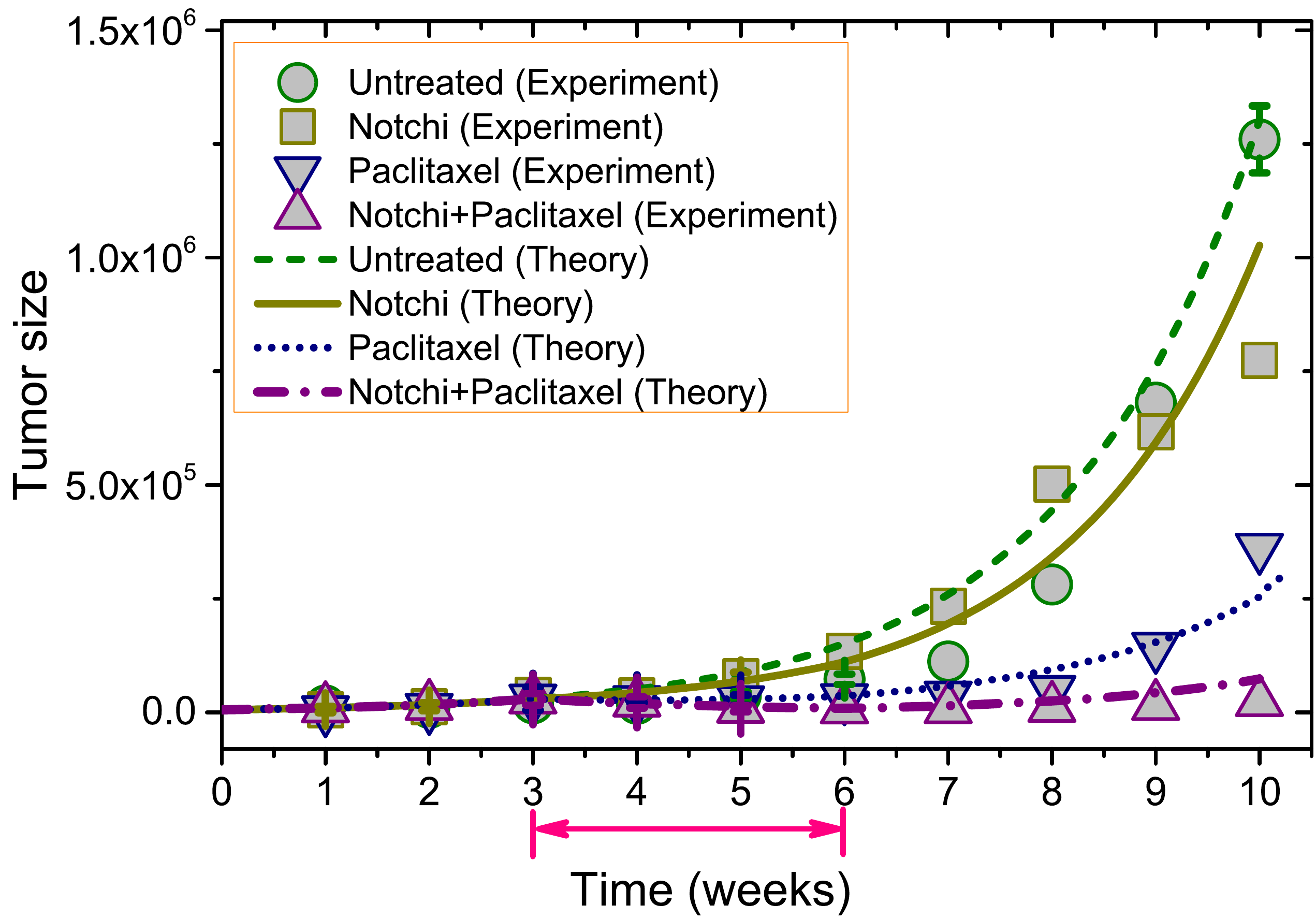}  } 
\caption{\label{drugresponse1} \textbf{The dynamics of tumor growth under different conditions.} The symbols represent results extracted from a recent experiment under four conditions\cite{jordan2016her2}: The green circle shows the growth of mammary xenografts generated from parental CTCs (a mixture of HER2+ and HER2- cells)  of breast cancer patients without any drugs. The dark yellow square and blue down triangle illustrate the tumor growth under treatment of Notchi inhibitor and Paclitaxel from the 3rd to the 6th week (indicated by the double-headed arrow), respectively.  The violet up triangle corresponds to the tumor growth under treatment of both drugs simultaneously in the same period of time.  The theoretical predictions for  tumor growth under  the four different cases are shown by the lines. The tumor is imaged using IVIS Lumina II. Its size is in the unit of the photon flux, which is proportional to the number of tumor cells. }
\end{figure}

\begin{figure}
{\includegraphics[width=1.0\textwidth] {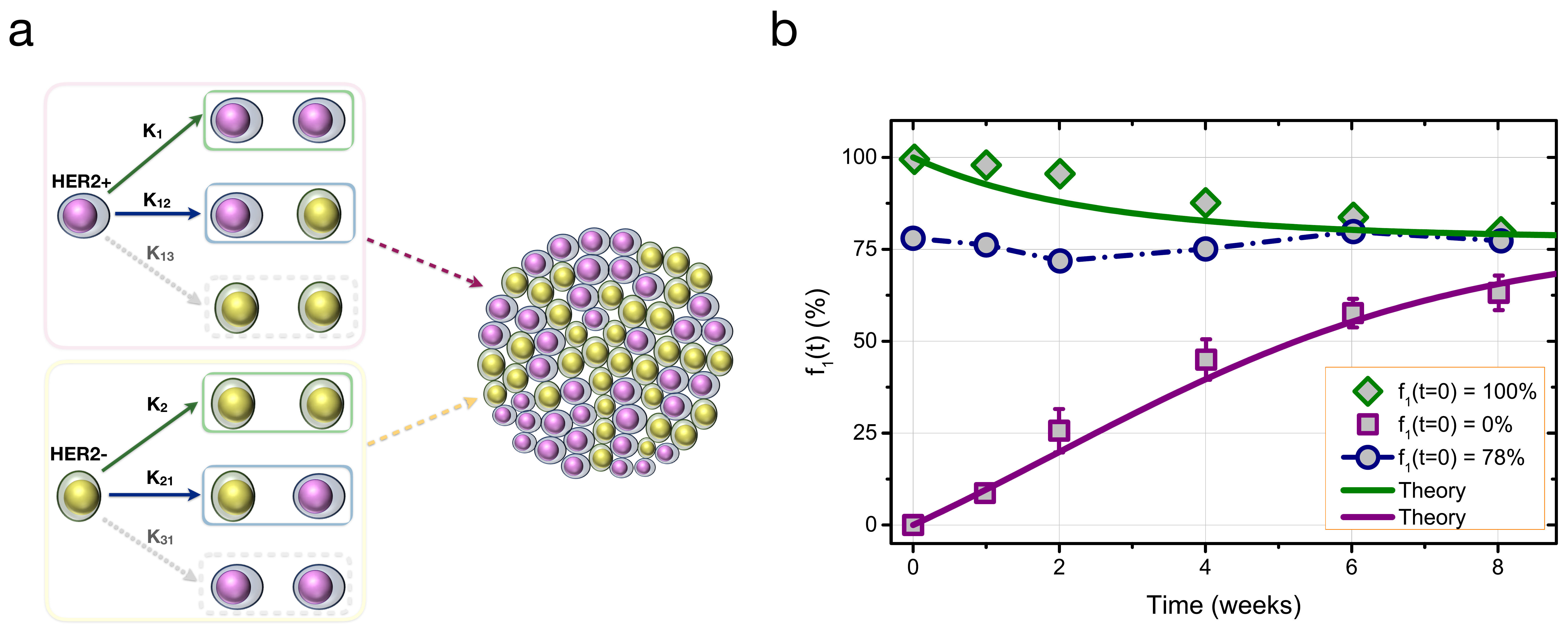}  } 
\caption{\label{figure2} \textbf{The dynamics of HER2+/HER2- cells.} ({\bf a}) Illustration of the ITH model for breast cancer.  Both HER2+ and HER2- breast circulating tumor cells (CTCs) may divide symmetrically, producing two identical HER2+ and HER2- cells with rates $K_{1}$ and $K_{2}$, respectively. They can also divide in an asymmetric manner by producing one HER2+ and one HER2- cell with rates $K_{12}$ and $K_{21}$. The two cell types could divide symmetrically but produce the other cell type (see the processes with rates of K13 and K31).  A  heterogeneous cell colony composed of both HER2+ and HER2- cells is established,  irrespective of the initial cell states.  ({\bf b}) Experimental data  for the fraction ($f_{1}(t)$) of HER2+ cells as a function of time for three  initial conditions: starting with HER2+ cells only (symbols in green), HER2- cells only (symbols in violet), and the parental cultured CTCs (symbols in navy). Theoretical predictions are shown by the solid lines.  The dash dotted line for the case of parental cultured CTCs is to guide the eye. }
\end{figure}

\textbf{Drug response in a heterogeneous breast cancer cell population:}
To set the stage for the mathematical model, we first summarize the results of experiments.  It is known that HER2+ cells appear in patients initially diagnosed with ER+/HER2- breast cancer during treatment\cite{arteaga2014erbb,houssami2011her2}. Although each cell subpopulation is sensitive to a specific drug, the heterogeneous tumor shows varying responses for distinct treatment protocols (see Fig.~\ref{drugresponse1} as an example).  The size of an untreated tumor increases rapidly (see the green circles), illustrating that a mixture of two cell types together has the ability to promote tumor growth. A clear response is noted when Paclitaxel (targeting HER2+ cells) is utilized, which results in reduced tumor growth (see the navy down triangles) during the treatment.  Surprisingly, the tumor continues to grow rapidly, with no obvious response, if treated by Notchi inhibitor (see the dark yellow squares). This is unexpected as the growth of HER2- cells (sensitive to Notchi inhibitor) is supposed to be inhibited by the drug.  Finally,  the combination therapy with both the drugs, Paclitaxel and Notchi inhibitor, administered to the tumors simultaneously effectively delays the tumor recurrence (see the violet up triangles). However, as both drugs have adverse toxic side effects on normal tissues\cite{kim2001vivo,tolcher2012phase}, the use of the two drugs simultaneously might not be advisable.  These observations suggest that instead of developing new efficacious drugs, more could be done to optimize the current treatment methods\cite{eisenhauer2021connecting}, which requires an understanding of the drug resistance mechanism, and evolutionary dynamics of each subpopulations  quantitatively. Here, we develop a theoretical model (see Fig.~\ref{figure2}a for the illustration of the model) to explain the occurrence of phenotypic heterogeneity in breast cancer, and explore diverse responses under different drug treatments (Fig.~\ref{drugresponse1}). 

\textbf{Phenotypic equilibrium in a heterogeneous cancer cell population:}  As mentioned above, it is found that HER2+ and HER2- breast cancer cells  transition from one  phenotype to another\cite{arteaga2014erbb}. To  demonstrate the observed cellular plasticity,  fluorescence-activated cell sorting (FACS)-purified HER2+ and HER2- cells were grown in culture for eight weeks independently in the experiments (see SI for more experimental details)\cite{jordan2016her2}. Surprisingly, HER2- (HER2+) cell,  naturally emerges from the initial HER2+ (HER2-) cell seeding within four weeks. The time course of the HER2+ cell fraction, $f_{1}(t)$, is shown in Fig.~\ref{figure2}b for different initial conditions. The fraction $f_{1}(t)$  decreases slowly, reaching a plateau with $f_{1} \approx 78\%$ after eight weeks of growth (see the green diamonds in Fig.~\ref{figure2}b) starting exclusively from HER2+ cells. On the other hand, $f_{1}(t)$ increases to $63\%$  (without reaching a plateau)  from zero  rapidly during the same time period, if the cell colony is seeded only from HER2- cells (see the violet squares in Fig.~\ref{figure2}b). Finally, the HER2+ cell faction, $f_{1}(t)$, almost does not change with time if the initial population is a mixture of both cell types derived from the parental cultured CTCs directly (see the navy circles in Fig.~\ref{figure2}b).  Therefore, a steady state level (with $f_{1} \approx 78\%$, the value in  the parental cultured CTCs) is established between the two different cell phenotypes at long times, irrespective of the initial cell fraction.
 
 
To understand the experimental findings summarized in Fig.~\ref{figure2}b,  we developed a mathematical model in which the cell plasticity is coupled to cell division,  as illustrated in Fig.~\ref{figure2}a (see SI for additional details). We first assume an equal rate $K_{12} = K_{21} \equiv K_{0}$ for the production of HER2- from HER2+ and vice versa.   We also neglected the symmetric division ($K_{13}$, $K_{31}$), one cell producing two identical daughter cells of the other type, because they rarely occur\cite{jordan2016her2}. We found that the two rates ($K_{12}$ and $K_{21}$) are small (see the following discussions), and it is not necessary to give different values in order to explain all the experimental results.   With these assumptions,  Eq.~(S3) in the SI can be simplified as,
\begin{equation}
 \frac{d f_{1}(t)}{dt} = (\Sigma-2K_{0})f_{1}(t)-\Sigma f_{1}(t)^2+K_{0} \ .\\
 \label{f1t}
\end{equation}
where $f_{1}(t)$ is the fraction of HER2+ cell in the whole population, and $\Sigma \equiv K_{1}-K_{2}$. Given the initial condition, $f_{1}(t=0)=0$, we find that $K_{0}=\frac{d f_{1}(t)}{dt}|_{t=0}$  from Eq.~(\ref{f1t}) directly. Therefore, the parameter value $K_{0} \approx 0.09$ per week is obtained using the first two data points from the experiments starting with only HER2- cells (see the violet squares in Fig.~\ref{figure2}b).  Finally, the value of $\Sigma$ can be calculated from Eq.~(S5) in the SI, which leads to $\Sigma \approx 0.3$ given the stable equilibrium condition ($f_{1}^{s} = 0.78$) found in the two cell populations in experiments (see Fig.~\ref{figure2}b). Hence, the time course of $f_{1}(t)$ can be calculated by solving Eq.~(\ref{f1t}), given any initial condition, $f_{1}(t=0)$, (see the two examples illustrated in Fig.~\ref{figure2}b by green and violet solid lines). Our theoretical predictions agree quantitatively with experiments, which is interesting considering that we only used two experimental data points.  We also found that the cell fraction conversion from HER2+ to HER2- is very slow, while the reverse process is rapid (see Fig.~S1 and discussion in the SI).


\textbf{Growth dynamics of cancer cell populations:}
 The CTCs of HER2+ have a higher proliferation rate compared to HER2-, as noted both in {\it in vitro} and {\it in vivo} experiments (see the green and blue symbols in Fig.~S2 in the SI). It is consistent with the predictions of our model, which shows that the rate difference, $\Sigma \equiv K_{1}-K_{2} \approx 0.3$, between the two cell types. Combined with the assumption that $K_{12} = K_{21} \equiv K_{0}$, it also explains both  the fast increase in $f_{1}(t)$ for the case when growth is initiated  from HER2- cells, and the slowly decay of $f_{1}(t)$ as initial condition is altered (Fig.~\ref{figure2}b).  The different dynamics of HER2+ cell is also associated with  it being a more aggressive phenotype, including increased invasiveness, angiogenesis and reduced survival\cite{dean2008her2,carey2006race}. 
 
To understand the growth dynamics of the cell populations as a function of  initial conditions (Fig.~S2) quantitatively, we need to determine either $K_{1}$ or  $K_{2}$. The other rate constant can be calculated using, $K_{1}-K_{2} \approx 0.3$. Using $K_{1}$ or  $K_{2}$, the growth dynamics can be derived directly from Eqs.~(S1) - (S2) in the SI with the condition $N(t) = N_{1}(t) + N_{2}(t)$ where $N_1(t)$ and $N_2(t)$ are the population sizes of the two cell types. The  model  quantitatively describes the growth behaviors of the tumor using only $K_{2} \approx 0.7$ (see the green and navy solid lines in  Fig.~S1 in the SI) as an unknown parameter. Note that $K_{2} \approx 0.7$ implies that $K_{1} \approx 1.0$. We can also predict the growth dynamics at different initial conditions, which could be tested in similar experiments. From the values of the rate constants, we would expect that the frequency for symmetric cell division (the two daughter cells are identical to the parent cell) is much higher than the asymmetric case for both the cell types ($K_{1}>K_{2} \gg K_{12}, K_{21}$). This prediction could be tested using single cell experiments.

\begin{figure}
{\includegraphics[width=0.80\textwidth] {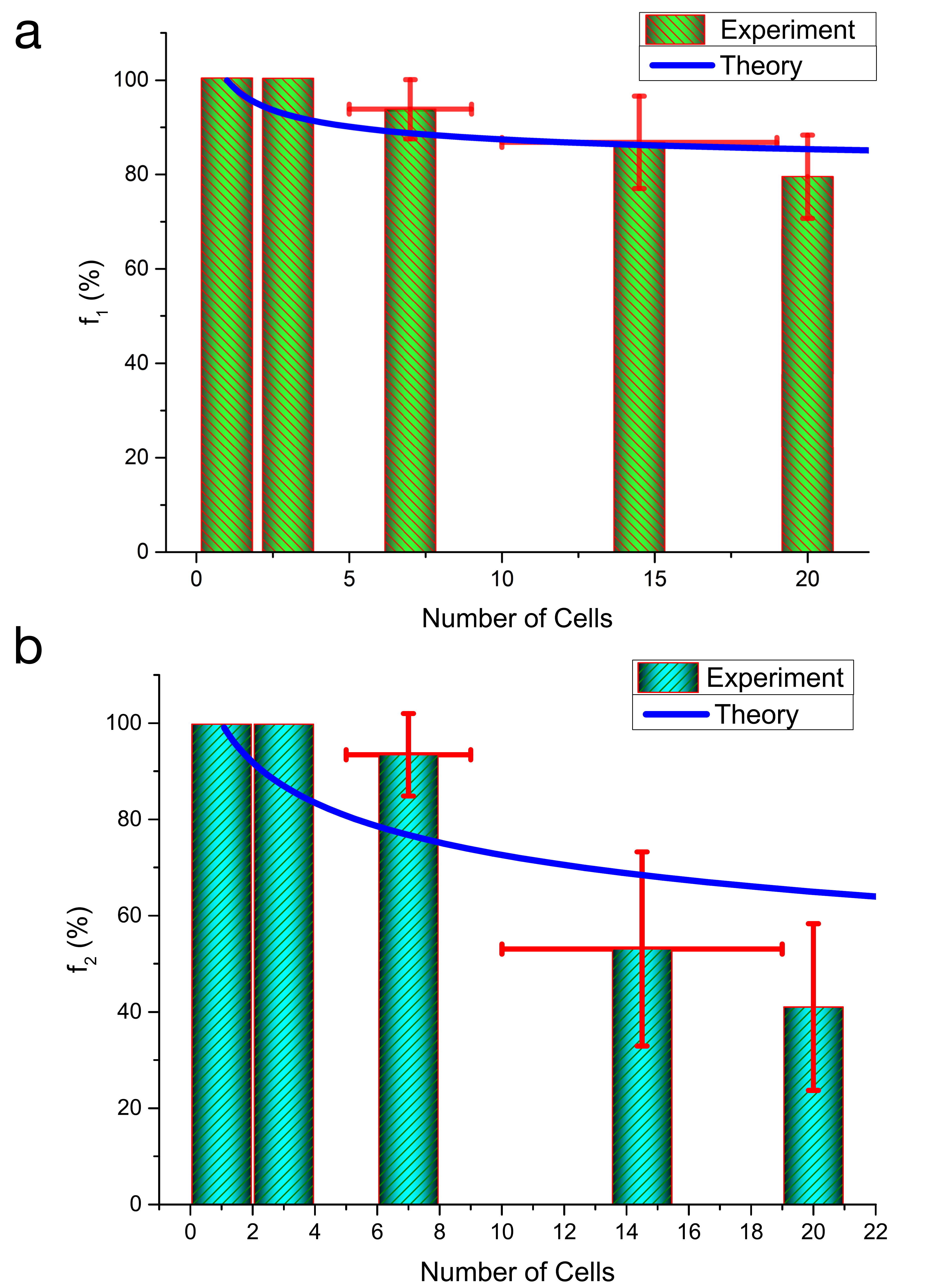}  } 
\caption{\label{figure3} \textbf{The interconversion between HER2+ and HER2- cell types.} ({\bf a}) The HER2+ cell fraction, $f_{1}$ (percentage), as a function of the total population size $N$ in a colony grown from a single HER2+ cell. ({\bf b}) The HER2- cell fraction, $f_{2}$ (percentage),  as a function of $N$ as the system develops from a single HER2- cell. The error bar in y-axis gives the standard variation, while the error bar in x-axis indicates the cell number range in which the cell fraction is calculated.  }
\end{figure}

\textbf{Cancer cell plasticity observed in single cell experiments:}
To further validate the model, we calculated the percentage of HER2+/HER2- cells as a function of the cell colony size starting from a single HER2+ or HER2- cell. The sizes of the cell colony have been measured in experiments (see the histograms  in Fig.~\ref{figure3})\cite{jordan2016her2}.  From Eqs.~(S1) - (S2) in the SI, we computed the HER2+ (HER2-) cell fraction, $f_{1} (f_{2})$, as a function of the cell colony size $N$ with the initial conditions, $N_{1}(t=0)=1$ and $N_{2}(t=0)=0$ ($N_{1}(t=0)=0$ and $N_{2}(t=0)=1$) using the same parameter values as given above.   Our theoretical results (see the solid line in Figs.~\ref{figure3}a and \ref{figure3}b)  are in good agreement with the experimental observations without adjusting any  parameter.  We also found that the HER2- cell fraction ($f_{1}$) decreases faster than the HER2+ cell fraction ($f_{2}$) as a function of the colony size ($N$), which is due to the higher symmetric division rate ($K_{1} > K_{2}$) of HER2+ cells.

Similarly, based on Eqs.~(S1)--(S2) in the SI or derived from the solid lines in Figs.~\ref{figure3}a and \ref{figure3}b directly, we calculated the cell colony size $N$, corresponding to the emergence of HER2+ cell starting from a single HER2- cell, and vice versa. The value of $N$ is around 5 and 8  obtained from our model for HER2+ and HER2- cells, respectively.  And the experimental values are found to be 5 to 9 cells, which agrees well with our theoretical predictions. Therefore, the model explains the experimental observation that one cell phenotype  can emerge from the other spontaneously after four cell divisions.

\textbf{Quantitative description for the drug responses of HER2+ and HER2- cell populations:} 
We next investigated the drug response in a heterogeneous population in (Fig.~\ref{drugresponse1}) using our model. Parameter values that are similar to the ones used to describe the experimental results {\it in vitro} are used but with minimal adjustments in order to capture the tumor growth observed in {\it in vivo} experiments. We rescaled the parameters $K_{1}$ and $K_{2}$ by a factor (2.06), which leads to $K_{\alpha}^{vivo} = K_{\alpha}/2.06$ with $\alpha =1$ or $2$ (see Table 1 in the SI). With these values, we found that  the tumor growth dynamics {\it in vivo} is recapitulated for the untreated tumor (see the green circles and dashed line in Fig.~\ref{drugresponse1}). 

HER2+ cells have a higher proliferation rate (see Fig.~S2 in the SI), and is sensitive to cytotoxic/oxidative stress (such as Paclitaxel treatment) while the HER2- cell shows a negligible response to Paclitaxel. On the other hand, Notch and DNA damage pathways are activated in the HER2- cell leading to  sensitivity to Notch inhibition. However, the HER2+ cells are resistant to  drugs for Notch inhibition\cite{jordan2016her2}.  To assess the influence of drugs on tumor  growth, we set the effective growth rate $K_{1}^{vivo}$ ($K_{2}^{vivo}$) of symmetric cell division to $-0.5$ (the negative sign mimics the higher death rate compared to the birth rate) when the drug, Paclitaxel (Notchi inhibitor), is utilized during treatment. We did not change the values of the asymmetric division rate constants, $K_{12}$ and $K_{21}$.     

Following the experimental protocol, we first let the tumor grow from a parental CTCs ($78\%$ of HER2+ and $22\%$ of HER2- cells) with an initial size taken at week one.  We then mimicked  drug treatment from the third week to the sixth week. Surprisingly,  our theory describes the growth dynamics of the heterogeneous tumor for different drug treatments well (see the different lines in Fig.~\ref{drugresponse1}). Our model successfully captured the inhibition of tumor growth under either Paclitaxel or the combination of Paclitaxel and Notchi inhibitor. Also the weak response of tumor under the treatment of Notchi inhibitor also emerges from our model naturally.

To understand the three distinct responses of the tumors to the drug treatments, shown in Fig.~\ref{drugresponse1}  further,  we computed the time dependence  of the tumor size in the first six weeks derived from our model with the treatment of either Notchi inhibitor or Paclitaxel  (see Figs.~S3a and S3b in the SI).  The tumor continues to grow rapidly without showing any clear response when  treated with Notchi inhibitor (see the symbols in navy in Fig.~S3a), inhibiting the growth of HER2- cells. Although unexpected, the observed response  can be explained from the cellular composition of the tumor. The  fraction of HER2+ cells is  high ($>70\%$) before  drug treatment, and it increases monotonically to even higher values ($\sim 90\%$) during treatment, as shown in Fig.~S3c in the SI.  Considering the proliferation rate of HER2+ cells is higher than HER2- cells, it is clear that tumor response under  Notchi inhibitor  only targets a minority of the tumor cell population and its reduction can be quickly replenished by the rapid growth of HER2+ (see the simple illustration in Fig.~S3e under the treatment of Notchi inhibitor).  Such a weak response is explained directly from the mean fitness, the growth rate $\omega = (K_{1}+K_{12})f_{1}+(K_{2}+K_{21})f_{2}$, landscape of the population, (see Fig.~S4 and detailed discussion in the SI). 



In contrast to the negligible effect of Notchi inhibitor to the progression of the  heterogeneous tumor, Paclitaxel treatment that targets the HER2+ cell leads to a clear reduction in the tumor size, and delays cancer recurrence (see Fig.~S3b in the SI). Such a response is due to the high fraction of the HER2+ cell in the tumor. It leads to the slowly growing of HER2- cells, which cannot compensate for the quick loss of HER2+ cells at the start of the treatment (see the rapid decay of HER2+ cell fraction in Fig.~S3d and Fig.~S3e for illustration).  However, the tumor recovers the fast growing phase in the fourth week (see Fig.~S3b) after the drug is used, corresponding to the time when  the fraction of HER2+ cell reaches around $0.5$ (derived from our model with $(0.5-K_{12})f_{1}(t) = (K_{2}^{vivo}+K_{21})f_{2}(t)$, and see also Fig.~S3d).  Once the fraction of HER2+  cells decreases to small values, the proliferation of resistant HER2- cells can compensate for the loss of HER2+ cells. Just as discussed above, such a response can also be seen directly from the fitness landscape of the population under treatment of Paclitaxel (see Fig.~S4 and detailed discussion in the SI).


The fraction of HER2+ cells quickly recovers to the value in the stationary state after drug removal (see Figs.~S3c and S3d), and the tumor grows aggressively again (see Fig.~\ref{drugresponse1} and Fig.~S3e for illustration).  Therefore, the progression of the  heterogeneous tumor cannot be controlled by a single drug, as demonstrated in the experiments, explained here quantitatively.

\textbf{Sequential treatment strategy:}
Our theory, and more importantly experiments,  show that the utilization of two drugs simultaneously could significantly delay the recurrence of tumors compared to the treatments using only a single drug of either type (see Fig.~\ref{drugresponse1}). However, the quantity of drugs used in the former protocol is much higher than in the latter case. Also, both drugs (Paclitaxel and Notchi inhibitor) have strong toxic side effects on normal tissues\cite{kim2001vivo,tolcher2012phase}. 
In the following, we consider a sequential treatment strategy with one drug followed by the treatment with the other, which would reduce the quantity of drugs used, and possibly  reduce the toxic side effects\cite{lin2020using}.

In the sequential treatment, there are two alternative methods depending on the order in which the drugs are administered.  We first let the tumor grows till the third week, and then apply the first drug, Notchi inhibitor (Paclitaxel), from the third to the sixth week followed by the utilization of the second drug, Paclitaxel (Notchi inhibitor), from the sixth to the ninth week. We used the same parameter values as taken in Fig.~\ref{drugresponse1}. Interestingly, we predict a dramatic difference between the responses of the tumors to the two treatment methods (see Fig~\ref{figure5}a). The tumor size shows no clear response to the treatment by Notchi inhibitor,  increasing rapidly  until Paclitaxel is used (see the circles in navy in Fig.~\ref{figure5}a and a schematic illustration in the upper panel of Fig.~\ref{figure5}c). From the phase trajectory (see the circles in Fig.~\ref{figure5}b), a rapid increase of HER2+ cell population ($N_{1}$) is found while HER2- cell population ($N_{2}$) decays slowly. In contrast, just as shown in Fig.~\ref{drugresponse1}, a clear delay is observed for the tumor growth when treated with Paclitaxel first followed by Notchi inhibitor  (see the diamonds in pink and navy in Fig.~\ref{figure5}a). Meanwhile, HER2+ and HER2- cell populations shrink rapidly during each drug treatment, as illustrated by the phase trajectory in Fig.~\ref{figure5}b (see the diamonds). It indicates the effectiveness of these two drugs.  In addition,  the tumor size is always much smaller in  the second protocol compared to the first,  reaching three fold difference in size (see the tumor size at the sixth week in Fig.~\ref{figure5}a). It follows that the order of drug administration greatly influences the treatment effects in the sequential treatment method, which is consistent with recent studies\cite{lin2020using,salvador2020cdk4}. We also illustrate the tumor response when treated with the two drugs simultaneously (see the pentagons in Fig.~\ref{figure5}a). A much better response is predicted compared to the first treatment method (see the circles in Fig.~\ref{figure5}a). However, the second approach shows a similar good response with a close tumor burden at the end of treatment (see the diamond and pentagon in Fig.~\ref{figure5}a). Hence, it is possible to find an optimal  strategy to obtain a similar treatment effect with attenuated side effect.

\textbf{Effect of duration of treatment:} In the previous sections, a futile treatment with rapid tumor growth is frequently found (see Fig.~\ref{drugresponse1} or the data in Figs.~\ref{figure5}a--\ref{figure5}b). We surmise that one drug should be removed at an appropriate time once it  produces no benefits. We studied the influence of treatment period length  ($\tau_d$)  on tumor responses. First, we investigated the sequential treatment by Notchi inhibitor followed by Paclitaxel for different $\tau_d$ values (see Fig.~\ref{figure6}a). The phase trajectories show that the variations in $N_{1}$, and $N_{2}$ and  their maximum values become smaller as $\tau_d$ is shortened. In addition, the response for each drug treatment is strengthened and the total tumor size (see the inset in Fig.~\ref{figure6}a) is always smaller for a smaller $\tau_d$. Therefore, a small $\tau_d$ should be used when such a treatment method is applied.

Next, we performed a similar analysis for the treatment with Paclitaxel  first, followed by Notchi inhibitor (see Figs.~\ref{figure6}b-\ref{figure6}c).  In contrast to the situation described above, the variations for $N_{1}$, $N_{2}$, and their responses to each drug treatment are similar even as $\tau_d$ varies. However, the total population size  (see the inset in Figs.~\ref{figure6}b-\ref{figure6}c) is smaller for the two-week treatment compared to  three and one-week treatment. We surmise that  instead of using one-week treatment for each drug, a two-week period would be a better choice in this treatment strategy.  Fig.~\ref{figure6} shows that the minimum values of $N_{1}^{min}$, $N_{2}^{min}$ (see Figs.~\ref{figure6}a and \ref{figure6}c) and  the total minimum tumor size $N^{min}$ (see the inset in Fig.~\ref{figure6}) at each treatment cycle  increases with time, irrespective of the value of $\tau_d$. This would result in uncontrolled tumor growth. In the following section, we will discuss potential approaches to control the tumor burden even if it cannot be fully eradicated.

\textbf{Control of tumor burden and Cellular plasticity leads to failure of treatments:} 
Despite the good response through certain treatment protocols as discussed above, tumor suppression is only transient, and the tumor recurs sooner or later due to drug resistance. Nevertheless, we can still seek, at least theoretically,  a stable tumor burden as a compromise, which is similar to the goals of adaptive therapy\cite{gatenby2009adaptive}. For the breast CTC consisting of HER2+ and HER2- cells, the model suggests that it is possible to control the tumor  maintained at a constant size (with relatively small variations, see Fig.~S5 and detailed discussion in the SI). Finally, we have learned from our calculations that the plasticity of breast cancer cells is one of the leading reasons for ITH, which in turn leads to drug resistance during therapy. We investigate how such a property influences the tumor response during treatment further. By varying the values of $K_{0} $ ($\equiv K_{12} = K_{21}$), we found that a strong transition between the two cell states can lead to total failure of treatments (see Fig.~S6a), while it is much easier to control the tumor burden as  the cellular plasticity is inhibited (see Figs.~S6b-S6c and more discussions in the SI). Therefore, theoretical models based on the tumor evolutionary process are likely to be useful in predicting the tumor progression, the clinical response, and possibly in  designing better strategies for cancer therapy\cite{li2020cooperation,anderson2008integrative,enriquez2016exploiting,malmi2018cell,sinha2020spatially}.

\section*{Discussion:}
We investigated the emergence of intratumor heterogeneity in breast cancer arising from cellular plasticity, which is embodied in the conversion between the HER2+ and HER2- phenotypes. In contrast to the unidirectional differentiation of normal stem cells\cite{wichterle2002directed,keller2005embryonic}, many cancer cells demonstrate a great degree of plasticity that results in reversible transitions between different phenotypes, leading to ITH without genetic mutations\cite{quintana2010phenotypic,gupta2011stochastic}. Such transitions are frequently observed in rapidly growing tumors, which is often neglected in theoretical models\cite{gupta2011stochastic}. Although some studies have recognized the need for taking a growing population, the models typically have many unknown parameters\cite{zhou2013population,zhou2014nonequilibrium}, which are hard to interpret. 

 By introducing a direct coupling between cell division and transition between phenotypes into a theoretical model, we provide a quantitative explanation for the emergence of a stable ITH,  a hallmark in HER-negative breast cancer patients. Our model accurately describes the evolution of different cancer cell fractions, and also the total tumor size observed in a recent single-cell experiment successfully. We predicted that the symmetric cell division appears more frequently compared to the asymmetric case for both types of cells found in breast CTCs. Without adjusting any parameter, our theoretical predictions for the cell fraction as a function of the cell colony size agrees extremely well with experimental results. The cell colony size (5$\sim$8 cells) calculated from our theory for the emergence of one cell phenotype from the other is in good agreement with the experimental observations (5$\sim$9 cells).

The asymmetric cell division has not been observed in the breast CTC experiment directly, although the experiment implies that cells of one phenotype produce daughters of the other phenotype\cite{jordan2016her2}. However, in a more recent experiment this was detected  in breast cancer\cite{granit2018regulation}. It was found that the newly formed cell doublet, after one cell division, can be the same cell type (symmetric division) or different  (asymmetric division, producing two daughter cells with one expressing the cytokeratin K14 while the other does not).  It is also possible that the state transition is not only coupled to cell division but can also appear through tumor microenvironment remodeling\cite{liotta2001microenvironment}. However, inclusion of these processes will add two more free parameters to our model, which is not needed to give the excellent agreement between theory and experiments.  In addition, such a state transition is not observed after cytokinesis was inhibited in  breast cancer experiment\cite{granit2018regulation}. Nevertheless,  our mathematical model could be extended to incorporate these possibilities should this be warranted in the future. 


Although the asymmetric cell division explains the bidirectional state transition, the underlying mechanism for such an asymmetric division is still unclear. In the experiments\cite{gupta2011stochastic,jordan2016her2,granit2018regulation}, the different states of cancer cells are mainly determined by the expression level of one or several proteins.  It is possible that these proteins (HER2, K14, etc.) are redistributed in the daughter cells unequally during cell division, which could be realized through a stochastic process or regulation of other proteins\cite{granit2018regulation,marantan2016stochastic,balazsi2011cellular}. 

The reversible phenotype transitions in cells have been found in many different types of cancers \cite{shaffer2017rare,van2017neuroblastoma,menard2001her2}, which not only lead to the development of drug resistance but also induce very complex drug responses, as discussed here.  Although each cell type is sensitive to one specific drug, the heterogeneous tumor derived from breast CTC shows an obvious response to Paclitaxel but not to Notchi inhibitor. Our model provides a quantitative explanation for the different time courses of the tumor under distinct treatments. The failure of the Notchi inhibitor, even at the initial treatment is due to its target, HER2- cell which is a minority in the heterogeneous cell population, and has a slower proliferation rate compared to the HER2+ cell.  Both experiments and our theory show  a significant delay of tumor recurrence under the combination treatment with two drugs applied to the tumor simultaneously.  We also predict that a sequential treatment strategy with Paclitaxel first,  followed by Notchi inhibitor (not in a reverse order of drugs) can show similar treatment effect as the one with two drugs used at the same time. In addition, the sequential treatment reduces the quantity of drugs administered each time, which can reduce the adverse effects in principle\cite{lin2020using}.   

One advantage of the mathematical model is that we can steer the evolutionary dynamics of each subpopulation by applying the right drug at the appropriate  time  to control the tumor burden.  This allows for a fuller exploration of the parameter space, which cannot be easily done in experiments.  Finally, we propose that patients could benefit from drugs which inhibit the  plasticity of the cancer cells\cite{granit2018regulation}. Taken together, our model could be applied to explore ITH found in other type of cancers\cite{shaffer2017rare,van2017neuroblastoma,menard2001her2,granit2018regulation}.   From the examples presented here and  similar successful studies, we expect that the physical and mathematical models may provide a quantitative understanding for the cancer progression and also stimulate new ideas in oncology research\cite{altrock2015mathematics,li2020cooperation,li2020imprints,bocci2019toward,gatenby2003mathematical}. We should emphasize that mathematical models  sharpen the questions surrounding the mechanisms of ITH, but real data from patients  are needed to understand the origins of ITH.

\section*{}
 \noindent 
 \textbf{Acknowledgements}
 
\noindent 
 We are grateful to Shaon Chakrabarti, Abdul N Malmi-Kakkada, and Sumit Sinha for discussions and comments on the manuscript. This work is supported by the National Science Foundation (PHY 17-08128), and the Collie-Welch Chair through the Welch Foundation (F-0019).
\noindent

\noindent 
\textbf{Author contributions}

\noindent 
X.L. and D.T. conceived and designed the project, and co-wrote the paper. X.L. performed the research.
\noindent 


\noindent 
\textbf{Competing financial interests}

\noindent 
The authors declare no competing financial interests.

 \bibliographystyle{vancouver}

\bibliography{heterogeneity17}

\clearpage
\clearpage
\floatsetup[figure]{style=plain,subcapbesideposition=top}
	
\clearpage
\clearpage
\floatsetup[figure]{style=plain,subcapbesideposition=top}


\clearpage
\clearpage
\floatsetup[figure]{style=plain,subcapbesideposition=top}


\clearpage
\floatsetup[figure]{style=plain,subcapbesideposition=top}
\begin{figure}
{\includegraphics[width=0.850\textwidth] {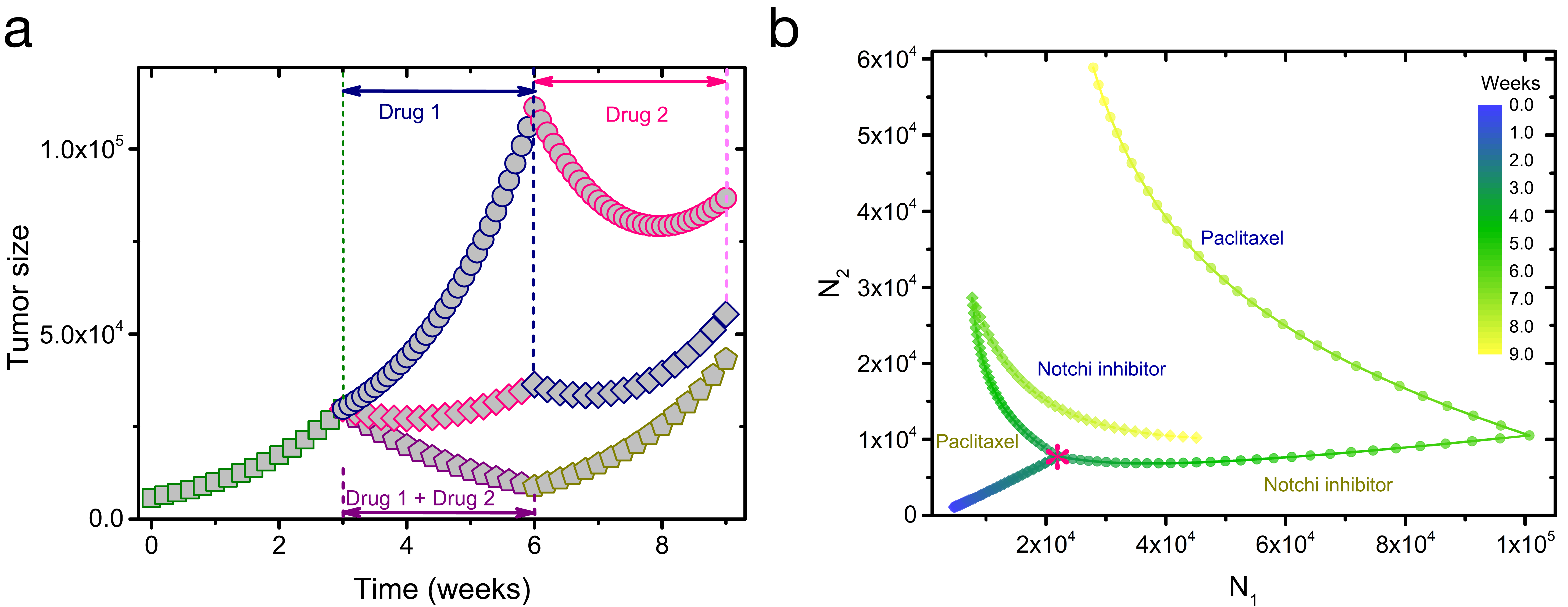} \label{figure5} } 
\caption{\label{figure5} {\bf Tumor response using a sequential protocol for two drugs.} ({\bf a}) Comparison of drug responses for tumors under different treatments. The green squares show tumor growth before treatment. The tumor under the treatment of  Notchi inhibitor first (navy), then Paclitaxel (pink) is indicated by the circles. The diamonds show the tumor growth under the reverse order of drug treatment, Paclitaxel first (pink), followed by Notchi inhibitor (navy). The pentagons demonstrate the treatment with both drugs administered simultaneously (violet color).  The pentagons in yellow show the tumor growth after the removal of all drugs.  The parameter values are the same as in Fig.~\ref{drugresponse1}. ({\bf b}) The phase trajectories for the two subpopulations,  HER2+ ($N_{1}$), HER2- ($N_{2}$),  under two sequential treatments considered in Fig.~\ref{figure5}a, respectively. The same symbols (circle and diamond) are used in ({\bf a}) and ({\bf b}). The initiation of the drug treatment is indicated by the red star and the trajectory color indicating the time is shown by the color bar. The drug name during each treatment period is also listed in the figure.     }
\end{figure}


\clearpage
\floatsetup[figure]{style=plain,subcapbesideposition=top}
\begin{figure}
{\includegraphics[width=1.0\textwidth] {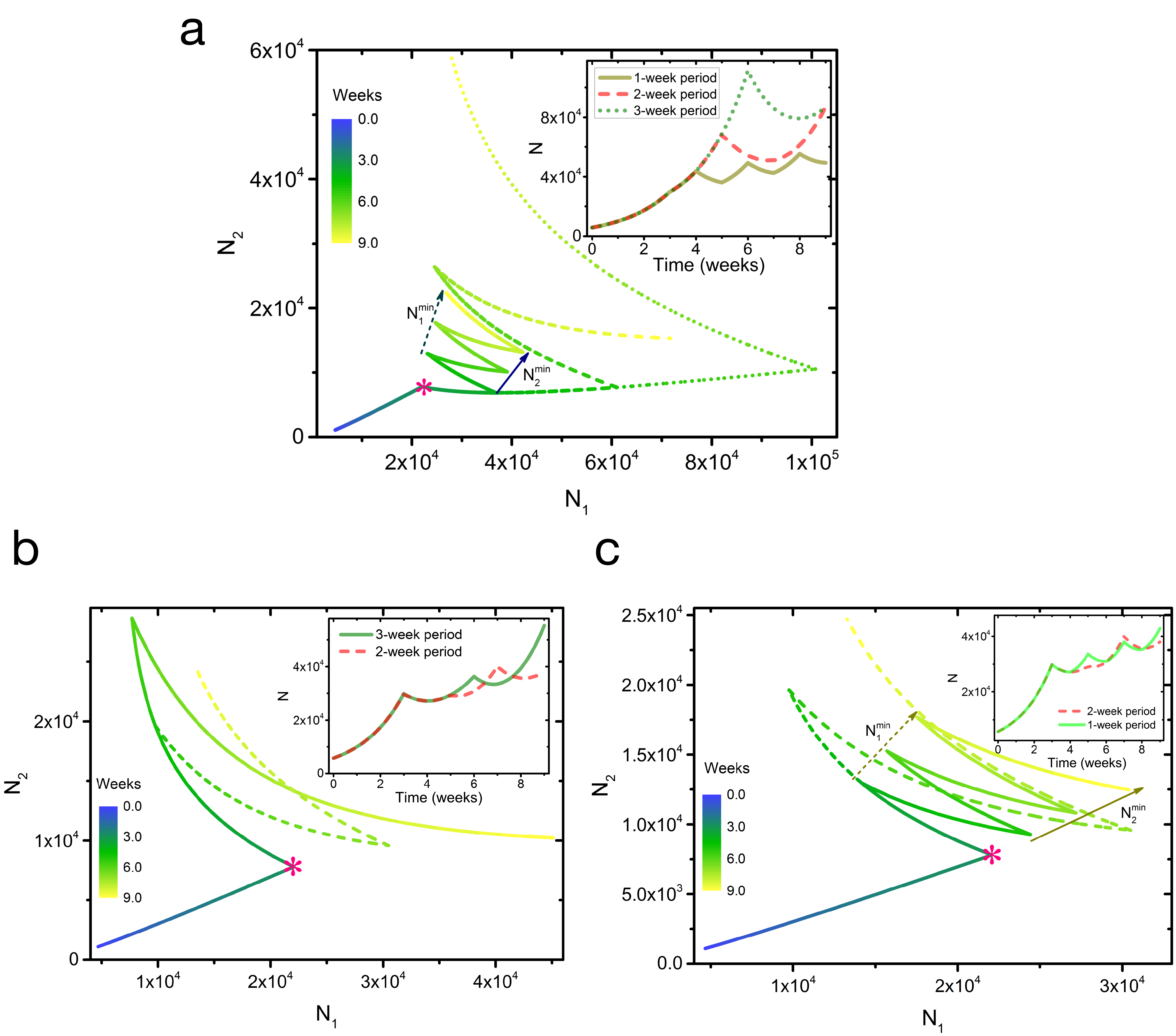} \label{figure6} } 
\caption{\label{figure6} {\bf  Phase trajectories for the two subpopulations as a function of duration  treatment duration.} ({\bf a})  Same as Fig.~\ref{figure5}b, treated by Paclitaxel first, followed by Notchi inhibitor,  except for the treatment period ($\tau_d$) for each drug being one (solid line), two (dashed line) and three week (dotted line), respectively. ({\bf b}) Same as Fig.~\ref{figure6}a but treated by Paclitaxel first, then Notchi inhibitor with a three (solid line), and two-week (dashed line) treatment period for each drug, respectively. ({\bf c}) Same as Fig.~\ref{figure6}b except for the treatment period for each drug being two (dashed line), and one week (solid line), respectively. The inset shows the total number  ($N = N_{1}+N_{2}$) of tumor cells as a function of time for different treatment periods.}
\end{figure}

\end{document}